\documentclass[conference]{IEEEtran}
\IEEEoverridecommandlockouts

\usepackage{cite}
\usepackage{amsmath,amssymb,amsfonts}
\usepackage{algorithmic}
\usepackage{graphicx}
\usepackage{textcomp}
\usepackage{xcolor}
\def\BibTeX{{\rm B\kern-.05em{\sc i\kern-.025em b}\kern-.08em
    T\kern-.1667em\lower.7ex\hbox{E}\kern-.125emX}}
\begin{document}

\title{Life-cycle Modeling and the Walking Behavior of the Pedestrian-Group as an Emergent Agent \\
{\footnotesize With empirical data on the cohesion of the group formation}
}


\author{
    \IEEEauthorblockN{Saleh Albeaik, Mohamad Alrished, and Faisal Alsallum}
    \IEEEauthorblockA{\textit{Center for Complex Engineering Systems} \\
    \textit{King Abdulaziz City for Science and Technology}\\
    Riyadh, Saudi Arabia}
}

\maketitle

\begin{abstract}
    This article investigates the pedestrian group as an emergent agent. The article explores empirical data to derive emergent agency and formation state spaces and outline recurring patterns of walking behavior. In this analysis, pedestrian trajectories extracted from surveillance videos are used along with manually annotated pedestrian group memberships. We conducted manual expert evaluation of observed groups, produced new manual annotations for relevant events pertaining to group behavior and extracted metrics relevant group formation. This information along with quantitative analysis was used to model the life-cycle and formation of the group agent. Those models give structure to expectations around walking behavior of groups; from pedestrian walking independently to the emergence of a collective intention where group members tended to maintain bounded distance between each other. Disturbances to this bounded distance often happened in association with changes in either their agency or their formation states. We summarized the patterns of behavior along with the sequences of state transitions into abstract patterns, which can aid in the development of more detailed group agents in simulation and in the design of engineering systems to interact with such groups. 
\end{abstract}


\section{Introduction}
    Autonomous multi-agent formations have been of interest to several scientific and engineering communities such as in aircraft and recently in drones \cite{gu2006design, du2017distributed}. Scientific communities studied animal behavior, such as the flocking of birds \cite{emlen1952flocking, chazelle2014convergence} as a reference to understand the nature of those formations and fundamental principals making them possible.

    The pedestrian group is a similar concept found in studies of human behavior, which proved to be of significant importance to the study of crowd behavior \cite{moussaid2010walking} and emergency evacuation \cite{von2017empirical, zhao2017analysis}. The concept naturally invited interdisciplinary research. A pedestrian group tends to be a social group first, and as such studied by social scientists \cite{stangor2015social, demerath2003social}. Those pedestrian groups tend to represent a significant portion of crowds inviting research from traffic research community \cite{chen2018social}. Physics community interested in deriving models that could be used to simulate such crowds gave the topic attention as well \cite{helbing1995social}. More recently, the engineering communities also found interest in this topic as they started to design socially aware robot systems expected to navigate through and interact with such crowds \cite{trautman2010unfreezing} and with the groups within \cite{katyal2022learning, alrished2025group}. 

    The walking behavior of a group received particular attention \cite{nicolas2023social, bandini2014towards}. This includes aspects such as the influence of group size on group walking speed and the shape the formation takes, as well as average group proxemics dispersion for different group sizes. Several aspects has been taken into account in scattered works to model and simulate the walking behavior of groups such as \cite{moussaid2010walking, qiu2010modeling, bandini2014agent}.

    Despite all the volumes of research in the topic and the variety of aspects studied and refined from bottom-up perspective, effort is still needed to organize available knowledge and fill in the gaps to help create unified frameworks describing the anatomy and construction of a group as an agent from a top-down perspective (discussed in next section). We believe such attempt could ease bringing available information in literature together and support construction of more accurate simulation models for the behavior of groups, as well as aid the design of engineering systems to integrate and interact with such groups.
    
    In this article as such, we start with an attempt on highlighting relevant information to the construction of such anatomy. We then delve into studying the life-cycle of the group as an emergent agent from empirical data and derive fundamental descriptive state spaces. We rely on datasets already accepted and utilized in literature for the analysis of crowd behavior capturing pedestrian trajectories with manual clustering of trajectories of pedestrian belonging to groups. We use this data to develop an agency and formation state spaces, to identify patterns of behavior such as boundedness of distances between group members, as well as disturbances to those bounds as group changes dynamically during the life-cycle of a group.  

\section{An Anatomy of a group}
    People with shared social ties tend to walk in groups and exhibit behavioral patterns such as shared direction and pace. While the literature is full of different definitions for what a pedestrian group is \cite{moussaid2010walking, zanlungo2014potential, stangor2015social}, here, from among those definitions, we assume that if a human inspector would identify them as a group based on their walking behavior, then they are a group.
    
    We thus start by focusing on a set of pedestrian who act as a pedestrian group. We assume each pedestrian can be represented with state variables representing information such as their position, velocity, target destination, etc. States relevant to social behavior can vary significantly and tend to be communicated using different modalities such as voice, appearance. To simplify this study, we focus our attention to information detectable from physical motion and its states. 

    \subsection{The group as an emergent agent}
    \label{sec:group_as_emergent_agent}

        The concept of an agent is explored in depth in philosophy and sociology as well as in computing and artificial intelligence literatures \cite{wooldridge1995intelligent}. For practical purposes, the concept is often used to refer to an entity capable of perceiving its environment, and taking actions to achieve its own goals and purposes. Each individual (human) within the crowd is an intelligent autonomous agent acting in complex ways. Multi-agent frameworks examine situations of complex interactions between those agents and highlight the emergence of collective behavior. 

\begin{figure*}[!htbp]
            \centering
            \includegraphics[width=\linewidth]{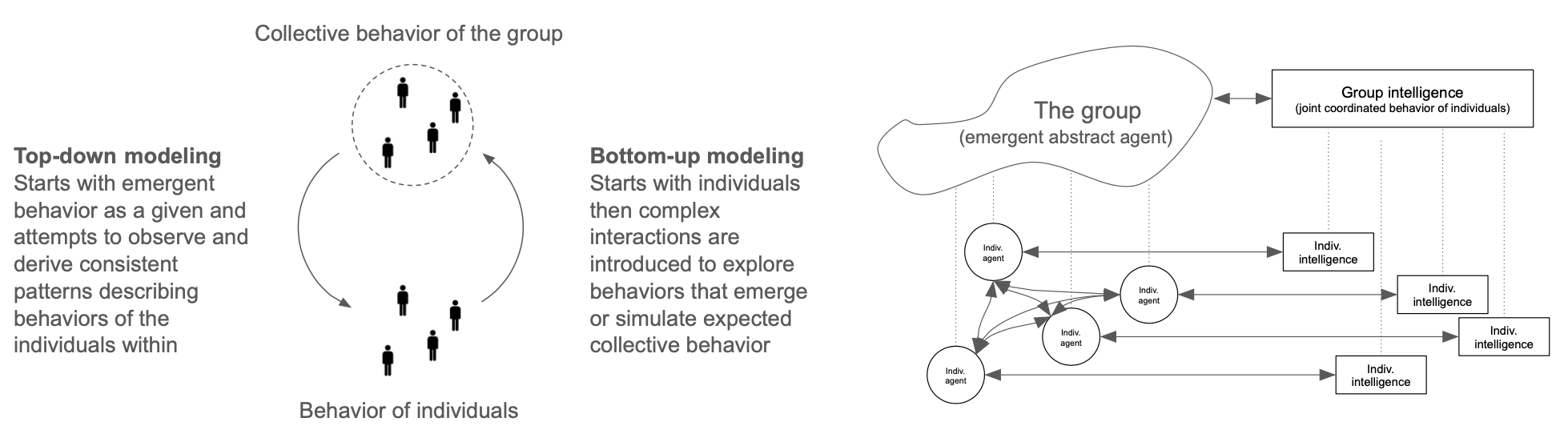}
            \caption{(Left) An illustration of top-down vs bottom-up modeling of emergent agency and patterns of emergent behavior. (Right) diagram representing individuals as agents and the group as an agent, the intelligence as a unit of each, and relationships between them.}
            \label{fig:group-agent-intelligence-structure}
        \end{figure*}

        This emergent collective behavior starts to become observable at the level of the pedestrian group. It has been modeled within agent-based frameworks such as \cite{bandini2014agent} (bottom-up perspective as illustrated in Figure~\ref{fig:group-agent-intelligence-structure}). On the other hand, collective agency theory postulates that the group itself can be seen as an agent \cite{misselhorn2015collective, hindriks2018collective, chang2006agent} and be studied as such (top-down perspective illustrated in Figure~\ref{fig:group-agent-intelligence-structure}). This group would be an emergent agent and can be described with terms such as an abstract-agent (or meta-agent) in reference to the fact that itself has no physical existence but is otherwise acknowledged through the collective intentionality of the group itself as well as the acknowledgment of other people to it. It can also be described using terms such as complex-agent in reference to its complex construction, complex interactions between the constituent agents, as well as complex emerging behavior through it. 

        The group as an agent would be said to have capacity for intelligent action, representing the effective intelligence of the constituent collective. This could be represented as in Figure~\ref{fig:group-agent-intelligence-structure}, where we explicitly declare the group as an abstract agent, and represent the intelligence of this abstract agent as emerging from the interaction of the intelligences of the constituent agents. 

        To examine the group as an agent, we explore concepts such as their cohesion as a unit (in this article), emergence of social norms around the group such as group proxemics and permeability as a reflection to level of social adherence to and acknowledgment of this unit (discussed in detail in \cite{albeaik2025proxemics}). In the next subsection, we discuss states relevant to this study and that are used within the subsequent quantitative analysis.

    \subsection{Structure and state space of the group as an emergent agent}
        Starting with the assumption that the group itself could be seen as an agent, here, we attempt to construct the structure that could be used to describe such a group and its walking behavior. A group could be described using composite state spaces that describe different aspects about the group, and influences its behavior. This include purely physical aspects such as:
        \begin{itemize}
            \item The set of members of the group.
            \item Physical position of each of those members.
            \item Physical motion state of each member.
        \end{itemize}

        This could be extended to include more abstract structures such as:
        \begin{itemize}
            \item Hierarchy between members within the group and group structure (such as sub-groups within the groups) \cite{qiu2010modeling, costa2010interpersonal, zanlungo2013walking, feng2014understanding}.
            \item Group walking formation and the shape it takes (such as a straight line, v-shape formation, etc) \cite{moussaid2010walking, schultz2013group, zanlungo2014potential}.
        \end{itemize}

        Moreover, the group as an intelligent agent can be said to have an intention of its own (the effective collective intention and decision of its members \cite{couzin2005effective}) and makes its own decisions. This could be seen through persistent objectives such as target destination towards which the group is collectively moving (or lack thereof), decisions to temporarily weaken the cohesion of the group or violate certain expectations, as well as aspects such as coherency and alignment of group members within the group itself \cite{reynolds1999steering, kiefer2013quantifying}. 

        In what follows, we attempt to tackle this challenge by starting to understand and model the life cycle of this group agent and study consistent patterns associated with the cohesion of their formation.

\section{Naturalistic study of patterns of behavior of pedestrian groups}
\label{sec:naturalistic-study}
    \subsection{The general observed setup}
        In this article, we focus on studying pedestrian groups as they walk through mostly open spaces and interact with other pedestrians in the scene; i.e., we focus on crowd traffic scenarios and study group behavior within. We approach our work as a naturalistic behavior study where we depend on video surveillance recordings of such crowds undisturbed. Such observational data has been documented and published in literature by several groups such as \cite{pellegrini2009you, bandini2014towards}. Those recordings are often preprocessed to generate pedestrian movement trajectories in addition to other information such as pedestrian group identification and labeling. Such preprocessed data is available through projects such as the OpenTraj project \cite{amirian2020opentraj}.

    \subsection{Datasets}
\begin{figure}[!htbp]
            \centering
            \includegraphics[width=\linewidth]{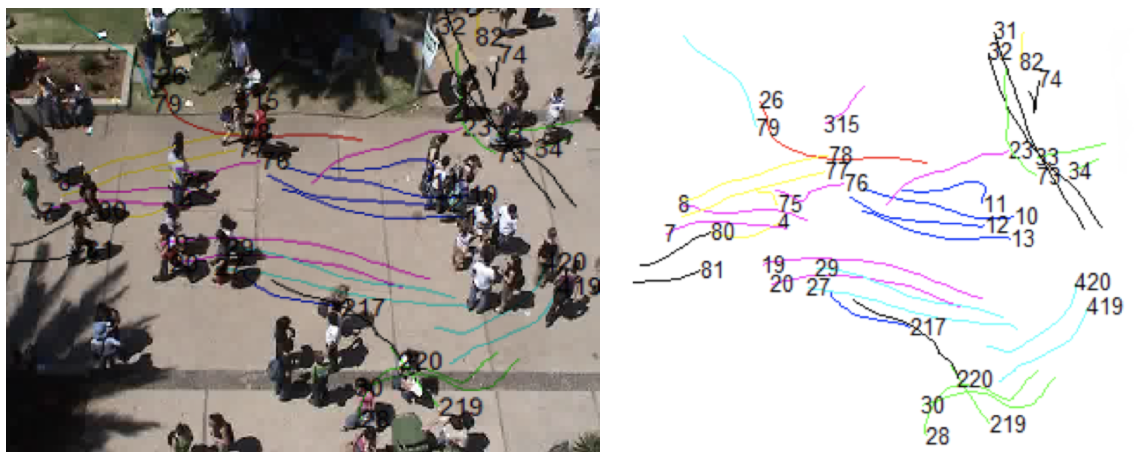}
            \caption{An image representing a general scene of observations with extracted pedestrian trajectories and group labeling (trajectory color). This is a sample from Student003 dataset \cite{lerner2007crowds}, with group annotation and illustration from \cite{solera2013structured}. This figure is borrowed from \cite{solera2013structured}.}
            \label{fig:dataset-example-image}
        \end{figure}

        In this article, we focus on studying datasets collected from several crowd environments such as students at university, urban street, airport, and other public spaces. We focus on datasets being used heavily for pedestrian behavior studies, manually annotated, and captures different scenarios that vary along relevant dimensions that influence behavior such as crowd density, traffic types and directions, and demographics. Specifically, we use the following datasets:
        \begin{itemize}
            \item ETH-Univ and ETH-Hotel datasets \cite{pellegrini2009you}.
            \item GVEII dataset \cite{bandini2014towards}.
            \item Student003 dataset \cite{lerner2007crowds}.
            \item Collective Motion Dataset (CMD) \cite{zhou2013measuring}.
        \end{itemize}
        Trajectories and group annotations used in this work were conducted and published by \cite{amirian2020opentraj} and \cite{solera2013structured, solera2015socially}. A sample from such observation setup, along with a sample of the extracted trajectories and group labeling is shown in Figure~\ref{fig:dataset-example-image}.

    \subsection{Preprocessing, annotation, and preliminary investigation}
        In this article, to conduct a focused study on pedestrian group behavior, we dissected each dataset into \emph{group observation} sets. That is, we extracted the groups from each dataset, identified the time frames where each group appeared in the dataset, and then extracted a copy of that part of the dataset to represent a group observation. A group observation as such is an annotated clip from a crowd scene where a specific group is observed from the time it enters the scene to the time it exits the scene. Other groups may appear in such a clip and would as such be considered as part of the background this ego group would interact with.

\begin{figure}[!htbp]
            \centering
            \includegraphics[width=\linewidth]{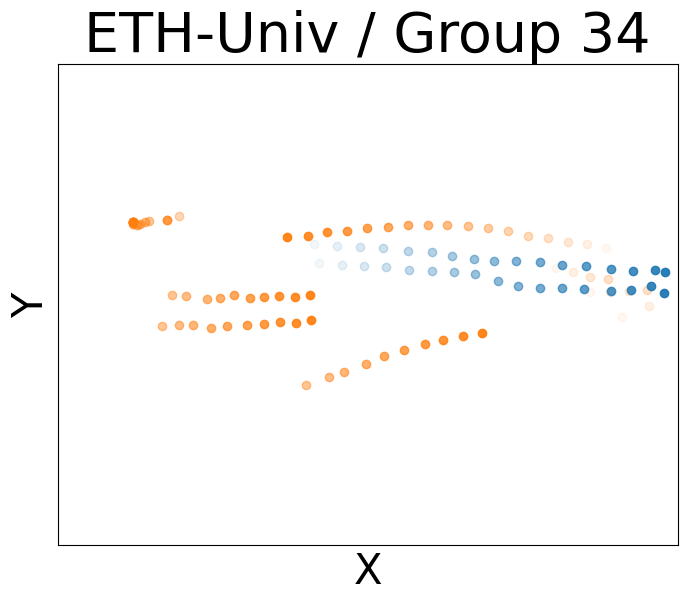}
            \caption{Sample trajectories extracted to represent a group observation (ego-group in blue) along with other pedestrian in the scene (in orange). Transparency highlights time.} 
            \label{fig:sample-group-observation}
        \end{figure}

        From this dissection, we generated a video clip for each group observation as shown in Figure~\ref{fig:sample-group-observation}, which we used for our preliminary manual expert examination. This examination was focused on generating initial insights into group behavior and identifiable patterns. We further used it to manually review published annotations of groups to identify any mislabeling (groups labeled as a group but label appears to be incorrect) that might affect our quantitative results. We further identified a set of recurring events affecting or disturbing what appears to be a qualitatively normal group behavior. This includes lack of direction, stationarity, significant member catch-up or split, and different types of social norm violations.

    \subsection{Preliminary hypothesis construction}
        We used the output from our preliminary annotation and investigation to start constructing a preliminary hypothesis on pedestrian group behavior, influencing factors, and potentially applicable state space description. In this article, the focus was primarily on the constituents and behavior of the group itself with emphasis on group agency, group cohesion, and group formation dynamics, which we found generally applicable across datasets and consistent across variations.

        Here, we focused primarily on two datasets, namely, ETH-Univ and GVEII capturing low and high density crowd scenarios. ETH-Univ further captures a wide range of group sizes up to groups of six members, while GVEII capturing a large set of group observations covering a wide spectrum of relevant cases to our study.

    \subsection{Iterative hypothesis refinement and quantitative analysis}
        Starting from the initial hypothesis, we conducted an iterative process of hypothesis refinement and quantitative investigation to identify and isolate most relevant factors, construct an emergent agency and formation cohesion state spaces, and derive recurring patterns of behavior.

        The authors of \cite{bandini2014towards} analyzed group proximity patterns (proxemics dispersion) and found that, on average, groups tend to maintain a relatively small distance between their members and that this distance grows proportional to group cardinality (number of group members, and we call it here group size). In their investigation, they measured dispersion as average distance between positions of group members and group center of mass.

        Here, we focus our investigation on maximum area covered and hypothesize that it is bounded, and identify conditions invalidating this boundedness assumption. To do so, we focus on group radius, defined as distance between group center of mass and physical position of group member farthest from center of mass.

        We also utilize an automated process to identify intrusions into the area the group is occupying indicating a loosening of the group formation. Here, we assume a conservative zone around the group represented by the convex hull covering group members, which has been used in literature for this purpose \cite{katyal2022learning}. 

        In the following sections, we present the resulting models of agent and formation state spaces, and present empirical observations and results from this analysis. 
        
\section{Modeling of the group agency and formation state spaces}
\label{sec:modeling-group-state-spaces}
    
\begin{figure*}[!htbp]
        \centering
        \includegraphics[width=\linewidth]{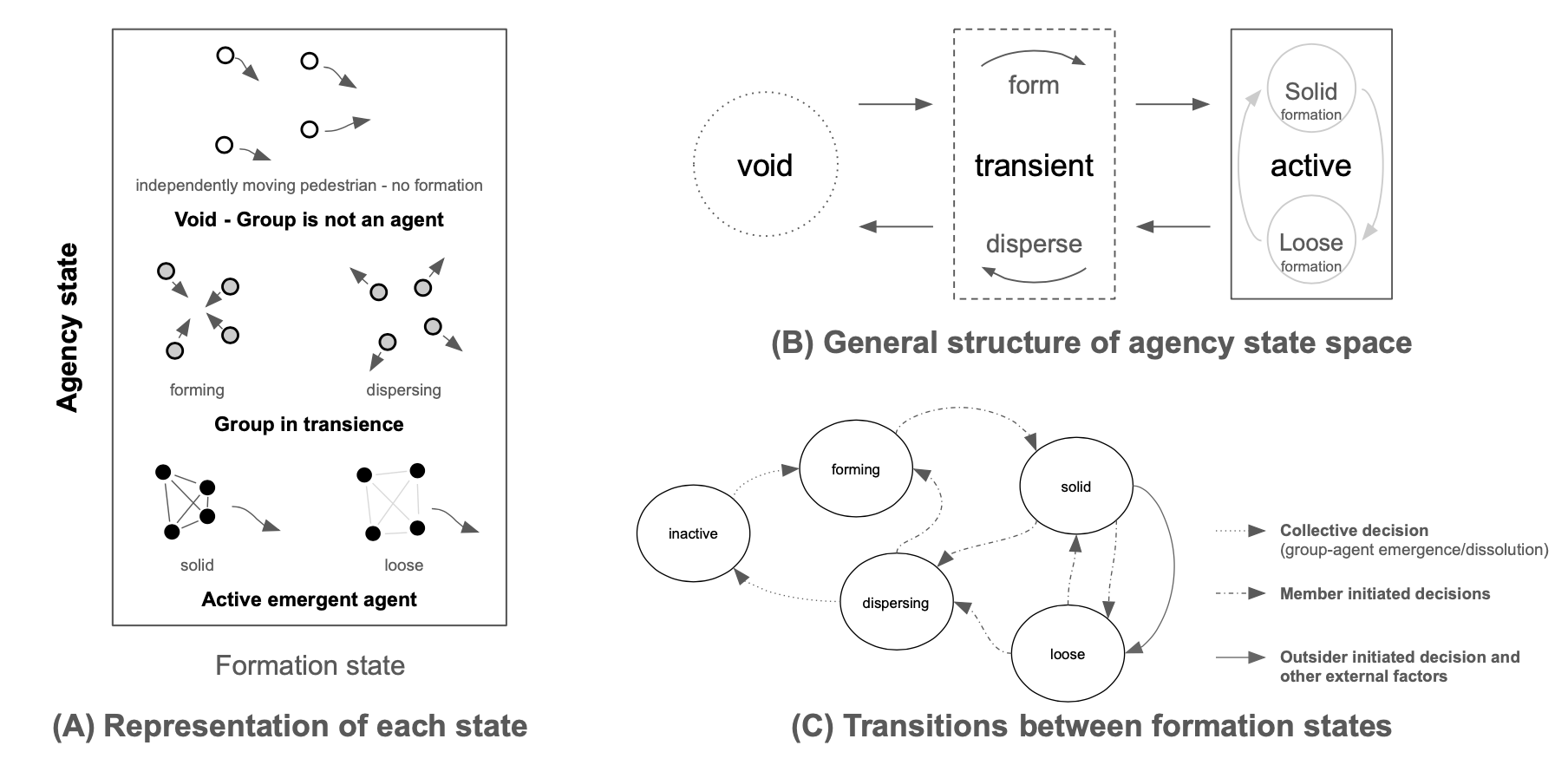}
        \caption{Group agency and formation state spaces. (A) An illustration for group members within each agency and formation state. (B) High level structure of group agency states and associated possible formation states. (C) Group formation states along with transitions and transition triggers.}
        \label{fig:combined-group-cohesion-state}
    \end{figure*}

    In this article, we attempt to extend this understanding of the behavior of a group beyond their cohesive state commonly studied. Based on our analysis and investigation discussed in the earlier sections, we propose that a pedestrian group as an agent can be described by an agency state variable and a formation cohesion state variable as illustrated in Figure~\ref{fig:combined-group-cohesion-state}. The agency state variable describes the state of the group overall while the formation cohesion state variable is specific to the condition of the formation they maintain as they move. 

    Next, we articulate and discuss each state variable and the associated state space, then discuss transitions within those state spaces.

    \subsection{The agency life-cycle state variable}
        Agent life-cycle models has been used in multi-agent systems and agent-based modeling to give structure to system design and implementation \cite{brazier2002supporting} with focus on triggers and transitions. In this article, we use agent life-cycle modeling to support the derivation of the group-agent model and associated consistent patterns of behavior within each state of the life-cycle. As discussed in Section~\ref{sec:group_as_emergent_agent}, the group can be seen as an emergent agent. Our observations have indicated that the group as an emergent agent is a dynamic condition, which changes during the time duration a group is observed. Here, we note three distinct states that observed groups experienced (or manifested) as summarized in Figure~\ref{fig:combined-group-cohesion-state}-(a).

        The following paragraphs highlight each state and associated behavioral patterns. The description makes references to the Social Force Model SFM \cite{helbing1995social}, an established model to simulate pedestrian walking behavior. Within SFM, pedestrian walk according to a set of forces such as goal attracting force, collision avoidance repulsive force, and within the extended model capturing group behavior \cite{moussaid2010walking}, an attracting group cohesion force. 

        \begin{itemize}
            \item \textbf{Void State:} In this state, the members of a group were observed to walk as independently moving pedestrian. Each member of this (yet to become or that were at some point a) group act independently and each with its own goal (target destination). In this state, there is no effective collective intention and they are generally undetectable as a group based on their observable behavior. When seen through SFM, group cohesion force would be described as weak or non existent (inactive), and that the active forces are forces to individual goals, and collision avoidance forces. 
            \item \textbf{Transient State:} In the transient state, the intention of the group is either becoming or disappearing. That is, an independently moving pedestrian would start to develop an acted intention to form a group and act like one, but the general collective coordinated behavior of a group might not be observable yet. In this case, for example, they would start to move closer together for example but not necessarily be moving towards a clear unified goal nor that their formation is fully developed yet. A group that is dispersing would go through the opposite process where the collective intention starts to disappear, the cohesion of the formation that maintained start to loosen and members start to walk away from each other. When seen through SFM, members of this potential group act in a transition such that the group cohesion force is either strengthening or weakening. When they are joining/forming the group, the dominating force tend to be that of group cohesion, while other forces tend to weaken or having weaker influence on the behavior of group members. 
            \item \textbf{Active State:} Once group members reach a stable group formation, their agency state could be described as active; i.e., they start to act collectively as an agent and could be seen/described as one. In this state, they tend to start to behave with a unified collective intentionality and goal, they start to exhibit collective behavior (as discussed in the following sections), and other people start to perceive them as such leading to emergence of social norms around the fully cohesive group (more detailed discussed in \cite{albeaik2025proxemics}). When seen through SFM, they tend to behave such that group cohesion forces and collision avoidance forces within the group are stable, and such that they act as a cohesive unit moving towards a shared target destination. 
        \end{itemize}
        
    \subsection{The formation state variable}
        While this variable can be highly intertwined and correlated with the group agency state variable in most discussions, we propose that those two variables should be seen as two distinct variables. The group agency state variable focus on group collective intention, the extent to which collective behavior is detectable and maintained by the group, the extent to which an outsider would identify the group as a group, etc. On the other hand, this formation state variable focuses on describing the formation the group maintains given they already have that collective intent. The states could be listed as conditional to the group agency variable as follows:

        \begin{itemize}
            \item Within Void group agency state, formation state are as follows:
                \begin{itemize}
                    \item Formation state is also \textbf{void.} In this state, there is no formation or collective behavior.
                \end{itemize}
            \item Within Transient group agency state, formation states are as follows:
                \begin{itemize}
                    \item Formation state \textbf{forming:} Group members are moving closer together. 
                    \item Formation state \textbf{dispersing:} Group members are moving away from each other.
                \end{itemize}
            \item Within Active group agency state (pedestrian groups discussed in literature are usually groups in this state), the group is moving as a collective unit. Formation states are as follows:
                \begin{itemize}
                    \item Formation state \textbf{solid or cohesive:} In this state, the group maintain a very cohesive unit with generally high coherency and alignment between members.
                    \item Formation state \textbf{loose:} This state can be similar to the solid state, however, the members might loosen up the formation, start to move slightly farther from each other, and might exhibit some level of independent behavior (lower coherency or alignment). However, the overall collective behavior and intentionality would still be intact, and their general goal and direction is still the dominant driver to their behavior. 
                \end{itemize}
        \end{itemize}

    \subsection{State transitions}

        After defining state variables and the different states each can take, we next define the relationship between these states and drivers to transition from one state to the next. Figure~\ref{fig:combined-group-cohesion-state}-(B) gives a high level structure to those states and transitions between them, as well as the structure of the formation state in relation to the agency state. Figure~\ref{fig:combined-group-cohesion-state}-(C) depicts the detailed micro states (effective state combining both agency state and formation state), transitions between states and associated triggers to each transition. 

        In discussing transitions between states, an important distinction can be made about the initiator or the trigger to that transition as well as the state the group will take afterwards. A group and its members generally make their own decisions to transition from one state to another, however, it is not too uncommon to observe situations where external factors drive or influence this decision (discussed in more details in \cite{albeaik2025proxemics}). For instance, an outsider to the group violates the social norm not to intrude into the group and that can drive the group to loosen its formation temporarily until the intrusion is over and it is possible for them to bring their group back to their solid cohesivity (solid to loose driven by external factors, with loose to solid transition temporarily blocked, and then loose to solid transition driven by group collective decision). Furthermore, certain constraints could also be noticed. For instance, a group in void state and group members far from each other would have to go through a transient (forming) state before they are able to enact a solid formation state. When we consider group hierarchy as well (we keep discussion of group hierarchy to minim in this article to focus on the core idea), sub-groups could be in different states. One commonly observed scenario is where a sub-group is cohesive, and one member or more are in catch-up or split state. Those members in cohesive state move cohesively to their goal, while the remaining members are walking towards the group itself to catch-up. 

        In the following section, we would explore empirical observations about these states, and highlight observed patterns to transitions between those states. 
    
\section{Empirical data and analysis results}

    This section examines empirical quantitative data of group observations. We focus primarily on analyzing group radius (discussed in Section~\ref{sec:naturalistic-study}) in relation to the state spaces (discussed in Section~\ref{sec:modeling-group-state-spaces}).

    In the following subsections, we start by investigating two primary datasets closely and discuss our observations and conclusions. We then use those observations and conclusions to derive a set of abstract patterns to describe and understand the general behavior of a group. We then conclude this section by presenting comprehensive results from a larger dataset selection and guide the reader through analyzing those datasets in the absence of manual labels that we used to guide our investigation of the two main datasets used initially. 
    
    \subsection{Investigation of group radius from empirical data}

        \begin{figure*}[!htbp]
            \centering
            \includegraphics[width=\linewidth]{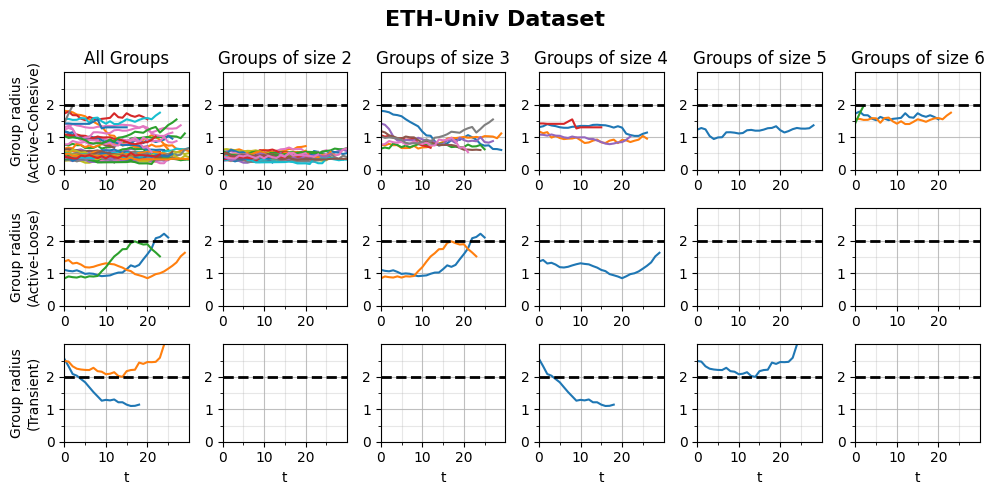}
            \caption{Group radius as a function of time measured for each group observed in ETH-Univ dataset. Radii for groups in different agency and formation states are isolated, as well as for each group size, to highlight patterns of behavior.} 
            \label{fig:group-radius-from-dataset-0}
        \end{figure*}

        \begin{figure*}[!htbp]
            \centering
            \includegraphics[width=\linewidth]{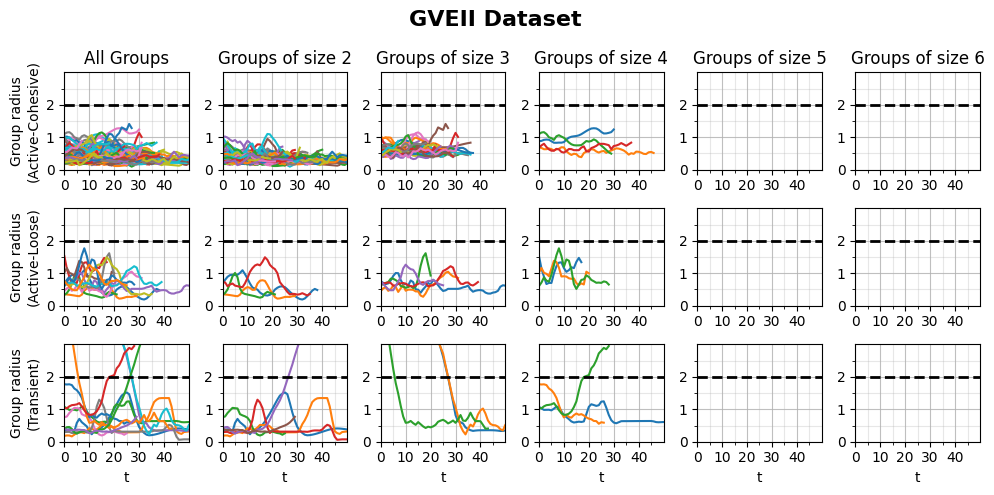}
            \caption{Group radius as a function of time measured for each group observed in GVEII dataset. Radii for groups in different agency and formation states are isolated, as well as for each group size, to highlight patterns of behavior.}
            \label{fig:group-radius-from-dataset-22}
        \end{figure*}
        
        In this subsection, we focus on ETH-Univ and GVEII datasets. Those are two of the richest datasets accessible to us, largest, most representative, and widely used in crowd walking behavior research. ETH-Univ is the only with samples representing all group sizes from groups of two to groups of six pedestrians. GVEII on the other hand has denser samples and captures a more diverse set of scenarios as would be noticed in the discussion below. 
    
        Here, we analyze group radius as a general indicator of group maintenance of their formation. We assume a group occupies an area descried by a circle, the centroid of the circle is the center of mass derived from the physical positions of each of the members of the group, and radius is the distance from this centroid to the farthest person in the group from this centroid. We plot this group radius, as seen in Figure~\ref{fig:group-radius-from-dataset-0} and Figure~\ref{fig:group-radius-from-dataset-22} as a function of time for the duration of a group observation. Each curve in those plots represent radius of one group for the duration they remained within the physical area observed (entry to exit from the scene). In these two figures, we eliminated groups that were identified as mislabeled group during our manual annotation process.

        In Figure~\ref{fig:group-radius-from-dataset-0} and Figure~\ref{fig:group-radius-from-dataset-22}, we split observed groups according to the state (Section~\ref{sec:modeling-group-state-spaces}) they are assumed to be in. The two figures are structured in a similar way. The first row represent groups in their cohesive/solid state, the second raw represents groups in their loose state, and the third row represents groups in their transient (forming or dispersing) state. Note that a group may experience and transition between different states during the period of observation. We thus flag occurrences of a state during the observation time. For simplicity, we order states according to the significance of impact; transient state is considered more dominant (in terms of impact on group radius) than a loose state, and a loose state is more dominant than a cohesive state. A group is assumed to be in their cohesive state, unless they experienced loosening of formation or a transient state. Group transience was manually labeled, while group loosening of formation was restricted to cases of outsider intrusion, which was detected algorithmically from trajectories. Data is split column wise per group cardinality (number of people in the group). 

        During our iterative modeling and quantitative analysis process, to maintain scientific rigor, we refrained from changing our initial manual annotation of the data. For instance, in retrospect, outliers can be observed in Figure~\ref{fig:group-radius-from-dataset-0}. Upon re-examination of the video of the scenario, some groups experienced state transitions that were fuzzy or harder to detect, or are caused by factors we did not test for (loosening of formation due to group passing through a door as opposed to loosening of formation due to outsider intrusion). 

        From Figure~\ref{fig:group-radius-from-dataset-0} and Figure~\ref{fig:group-radius-from-dataset-22}, it can be observed that groups in their cohesive state generally maintain a practically constant and bounded radius for the duration of observation. This observation seems consistent across the two datasets, and across all group sizes observed. 

        Group radius also remained fairly small and bounded for groups that experienced loosening of formation. However, they experienced episodes where their group radius increased relatively. On the other hand, group radius was not bounded for groups that experienced transiency. Further, groups that experienced group transiency but not dissolution of the group, experienced radius disturbances of larger duration and magnitude compared to groups in their loose formation state.

    \subsection{Derivation of group radius and associated state transition patterns}

        \begin{figure*}[!htbp]
            \centering
            \includegraphics[width=\linewidth]{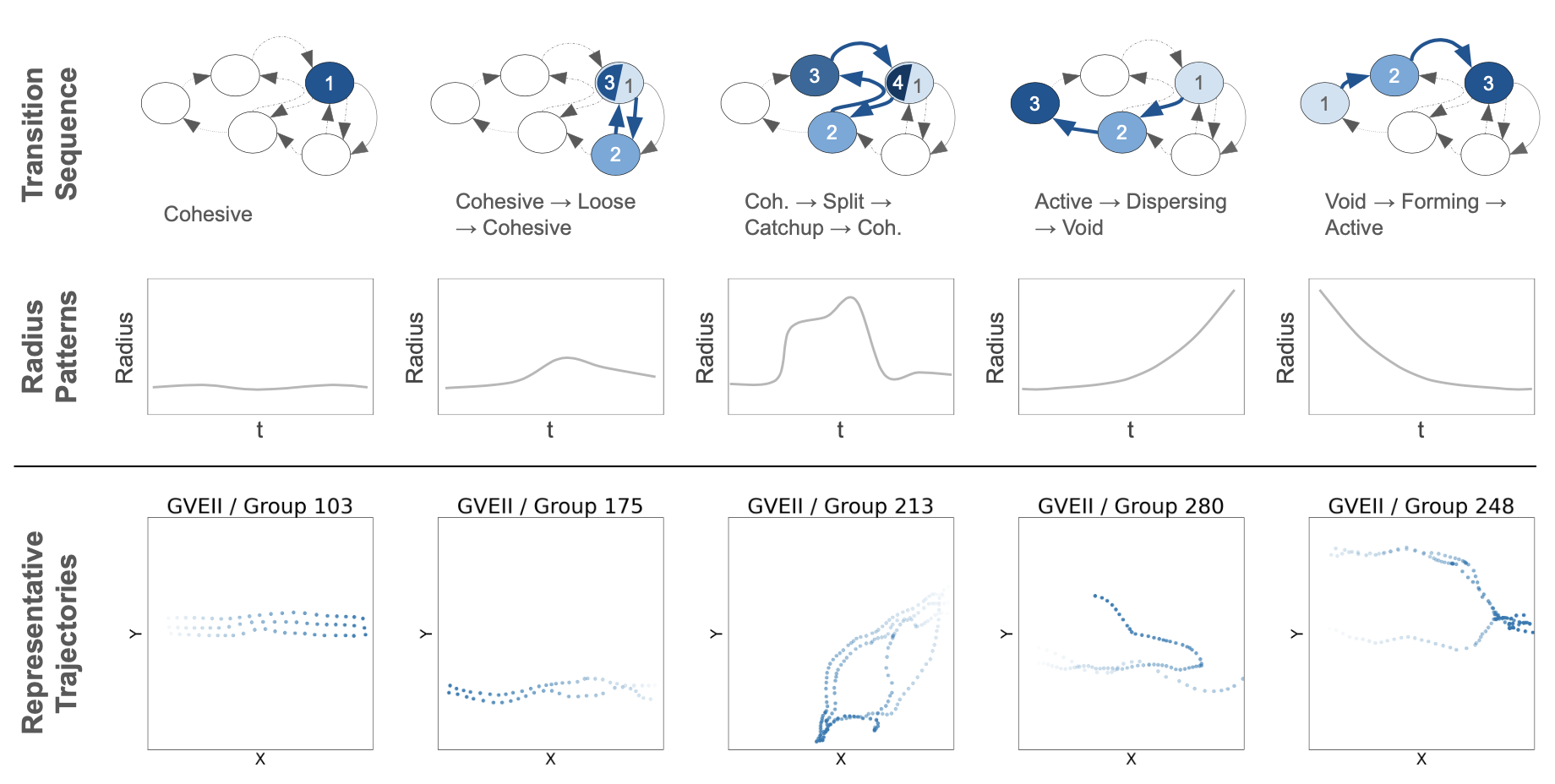}
            \caption{State and state-transition patterns observed in datasets. The first two rows illustrate the patterns we derive from observed behaviors and the last row presents representative group observation for each of the state and state transition sequence patterns.}
            \label{fig:state-transition-patterns}
        \end{figure*}

        In this section, we attempt to summarize the different patterns we observed in our data. We dissect the patterns to state transition sequences and associated group radius patterns, and we present them here along with trajectories from a representative group observation as shown in Figure~\ref{fig:state-transition-patterns}. For illustration purposes, we hid other pedestrian from the scene in those drawings.

        \begin{itemize}
            \item \textbf{Cohesive:} Those are groups that were in their cohesive state and remained cohesive for the duration of observation. Group radius was observed to be small and practically constant.
    
            \item \textbf{Cohesive $\longrightarrow$ loose $\longrightarrow$ cohesive:} Those are groups that were in their cohesive state but experienced temporary loosening of formation. Group radius was observed to be small initially when they were in their cohesive state, then experienced a disturbance in radius that was bounded and small in duration or magnitude and possibly larger in the other.
    
            \item \textbf{Cohesive $\longrightarrow$ dispersing $\longrightarrow$ forming $\longrightarrow$ cohesive:} Those are groups that started cohesive but decided to split temporarily and then reconvene. Radius of those groups started small and then grow for an extended period and magnitude compared to the loose state scenarios. During this time, they ceased to act as a group and would be harder to detect as a group should observation started from that moment.
    
            \item \textbf{Cohesive $\longrightarrow$ dispersing $\longrightarrow$ void:} Those are the groups that were detected as such initially but then decided to dissolve their groups and each member went their way. In this case, initial radius was small but then grew unboundedly until they are no longer a group.
    
            \item \textbf{Void $\longrightarrow$ forming $\longrightarrow$ cohesive:} Those are the groups that were detected as groups because they ended up acting like one at some point during the observation time, but not at the beginning. In those cases, independently moving pedestrian potentially coming from different directions convened to form a group. Once identified as a group, they would be considered to have unboundedly large radius at the beginning, but then that radius shrinks during their forming state until it stabilizes to a small radius when they reach their cohesive state.
        \end{itemize}

    \subsection{Examination of more comprehensive datasets}
        \begin{figure*}[!htbp]
            \centering
            \includegraphics[width=0.45\linewidth]{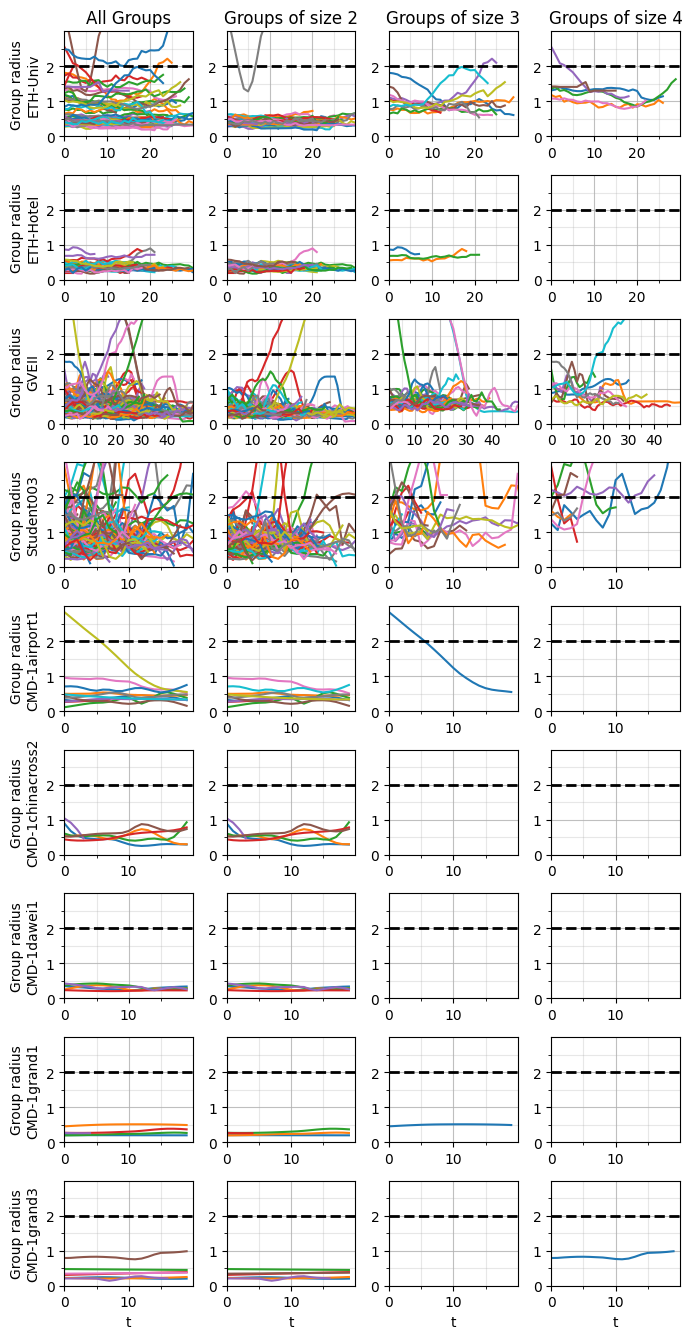}
            \includegraphics[width=0.45\linewidth]{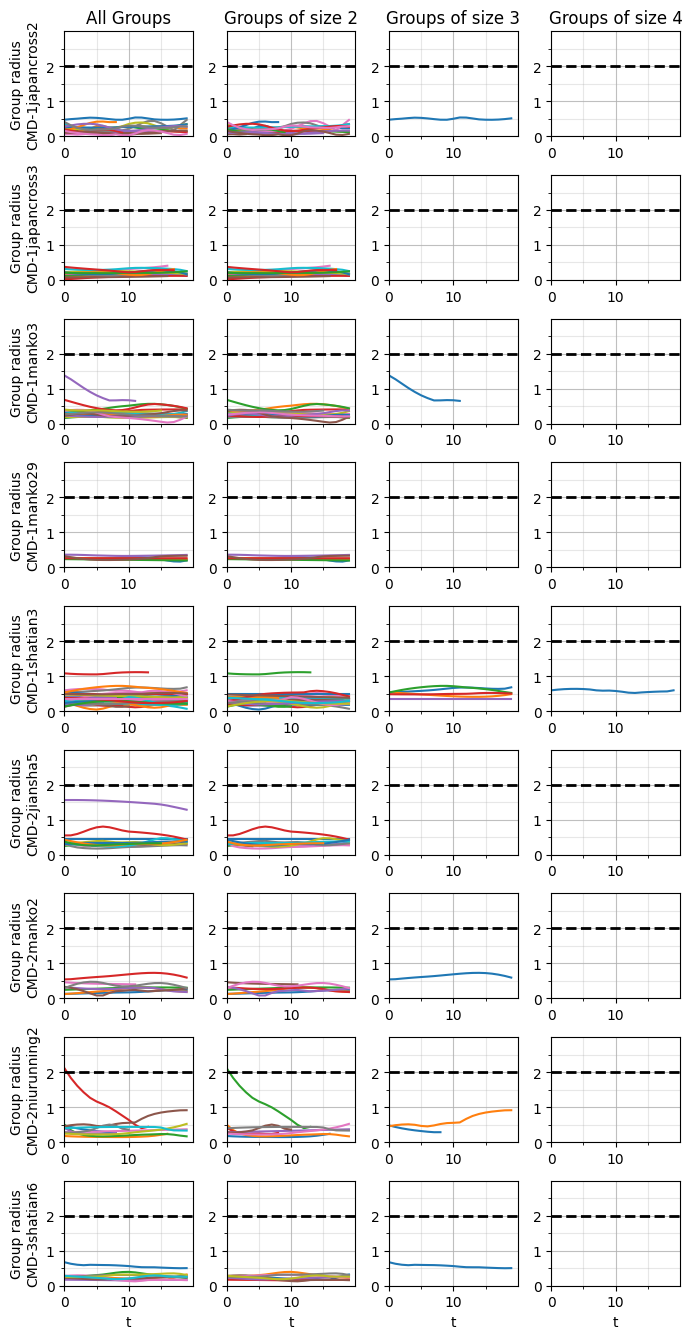}
            \caption{Group radius as a function of time measured for each observed group. This figure covers the four dataset collections discussed in Section~\ref{sec:naturalistic-study}.}
            \label{fig:group-radius-all-datasets}
        \end{figure*}

        In this section, we attempt to provide some insight into generalizability and replicability of those insights into other datasets. However, due to limited workforce for manual annotation, we present the remaining datasets from high level perspective as a reference for our readers in Figure~\ref{fig:group-radius-all-datasets}. We assume that the insights extracted in the earlier sections can be used to derive simple predictors to states and state transitions, which can be used to automate annotation for further investigation of those unlabeled sets. However, doing so would render this validation step unscientific and thus we refrain from doing so. 
        
        Figure~\ref{fig:group-radius-all-datasets} presents group radius for all groups in all datasets with group labeling available to us. We eliminated datasets from source that contained only five or less group observations. For presentation compactness, we also eliminated larger groups (groups of size five and six) as they are not frequently observed in data (only appear in ETH-Univ, which we presented above already, and a single instant of a group of size five in another of the datasets). In those plots, we present all data without applying any further filters. For instance, some groups that were eliminated in the subsections above due to mislabeling are retained here for completeness. 

        Through these plots, we point our reader to notice the general patterns we pointed out earlier in this article. This includes general boundedness and stability of group radius, and where this general observation fails, paying attention to recurrences of the abstract pasterns summarized above for groups experiencing changes in group state.

\section{Conclusions}
    This article investigated group agency and group formation through empirical data. The article focused on deriving a group agency state space and a group formation state space and studying associated observed patterns in group radius. We observed general boundedness and stability of radius for groups in their cohesive state, and identified general patterns associated with group transition sequences where the general boundedness and stability fails. In other articles, we show that those distinct agency and formation states have social behavior implications and consequences beyond the physical boundedness of distances between members of the group presented here.

\bibliographystyle{IEEEtran}
\bibliography{references}

\end{document}